# DESIGN OF SOFT VITERBI ALGORITHM DECODER ENHANCED WITH NON-TRANSMITTABLE CODEWORDS FOR STORAGE MEDIA

Kilavo Hassan[1], Kisangiri Michael[1], and Salehe I. Mrutu[2]

[1]Nelson Mandela African Institution of Science and Technology, School of Computational and Communication Science and Engineering, P. o. box 449 Arusha, Tanzania.
[2]University of Dodoma, College of Informatics and Virtual Education, P. o. box 490 Dodoma, Tanzania

*ABSTRACT*

*Viterbi Algorithm Decoder Enhanced with Non-transmittable Codewords is one of the best decoding algorithm which effectively improves forward error correction performance. HoweverViterbi decoder enhanced with NTCs is not yet designed to work in storage media devices. Currently Reed Solomon (RS) Algorithm is almost the dominant algorithm used in correcting error in storage media. Conversely, recent studies show that there still exist low reliability of data in storage media while the demand for storage media increases drastically. This study proposes a design of the Soft Viterbi Algorithm decoder enhanced with Non-transmittable Codewords (SVAD-NTCs) to be used in storage media for error correction. Matlab simulation was used in this design in order to investigate behavior and effectiveness of SVAD-NTCs in correcting errors in data retrieving from storage media.Sample data of one million bits are randomly generated, Additive White Gaussian Noise (AWGN) was used as data distortion model and Binary Phase-Shift Keying (BPSK) was applied for simulation modulation. Results show that,behaviors of SVAD-NTC performance increase as you increase the NTCs, but beyond 6NTCs there is no significant change and SVAD-NTCs design drastically reduce the total residual error from 216,878 of Reed Solomon to 23,900.*

## 1. INTRODUCTION

Storage media devices are highly demanded in almost every section including individual users, personal computers, mobile devices, data servers, cloud computing etc. As information technology advances, the demand for data storage media increases drastically[1, 2].The storage media can be categorized into three different models namely, directly attached storage, network attached storage and storage area network[3].Essentially, reliability of storage media is very important to minimize data loss.

The data storage industry faces challenges including devices limited lifetime, reliability, and failure. Yet, most data storage industries do not research on improving reliable and fault tolerant devices. They rather work hard in improving backup and recovery systems[1, 3].Among all types of storage media, Hard disk drives seem to have higher reliability or life span of about 3to 5years. However, in serious business hard disk drives are tied in Redundancy Array of Independent Disks (RAID) systems with higher frequencies of backups which indicate low trust of users on the devices and other possible environmental catastrophic events. The paper by Microsoft researcher in 2010 [4]show that the majority hard disks are replaced and found to have fault. It was reported that about 78% of all reported hardware replaced in Microsoft data centers are hard disks[4, 5].

"An analysis yields the following results. 70% of all server failures is due to hard disks, 6% due to RAID controller 5% due to memory and the rest (18%) due to other factors. Thus, hard disks are not only the most replaced component, but also are the most dominant reason behind server failure" [4].The stored data reliability can be increased by improving error correction or recovery mechanism in storage media[1].

The introduction of locked convolutional encoder to encode data and Viterbi Algorithm decoder enhanced with NTCs in Forward Error Correction (FEC) invites researches of the technique in storage media to enhance data reliability [6, 7].This new technique reduces the computation complexity while maintaining superiority in performance[8]. Computation complexity was one of the drawbacks of the application of convolutional codes in storage media. Currently Reed Solomon is almost the dominant algorithm used in storage media error correction, However storage media failure cases are still reported rampantly. Therefore, researchers in this paper propose a design of error correction mechanism (locked Convolutional encoder with Enhanced Viterbi Algorithm decoder) that can be applied in storage media to enhance reliability. The rest of this work is organized as follows: Section 2, designing of the proposed model; Section 3, simulation of the designed algorithm; Section 4, discussion and results, and Section 5 Conclusion and further recommendations.

## 2. DESIGN OF PROPOSED MODEL

The proposed model involves the following sub processes; data encoding using locked convolutional encoder on the data writing to the storage media and Soft Viterbi Algorithm Decoder Enhanced with NTCs during data reading process.

### 2.1. Locked Convolutional Encoder

Convolutional codes were first introduced by Elias in 1955[9, 10]. Since then they have been very popular in practical application. Convolutional codes work with data sequentially[11, 12]. Likewise, Convolutional codes have more powerful correcting capabilities than the block codes[13]. The codes are not only sometimes superior to block codes but also relatively simple to decode[14]. Essentially, Convolutional codes contain memory (m) and its encoding process is dependent on the current and previous message input. It has three parameter n, k and v, where n is the code words length, k is the message length and v is the constraint length which is defined as the number of previous messages involved, m plus 1, where m is the memory[15].

Convolutional code can be decoded by either sequential decoding or Viterbi decoding[16]. Convolutional encoding and Viterbi decoding are one of the powerful forward error correction techniques [12, 17, 18]. This design will focus on the Viterbi decoding because it is one of the powerful method for error correction and also due to the introduction of NTCs on Viterbi decoder. Viterbi algorithm can be a computer intensive kernel in Hidden Markov model based sequence alignment application [19-21].

To be able to use the technique of Non-Transmittable Codewords (NTCs) in data retrieving process from the storage media we need to use locked convolutional Encoder when writing data into storage media. Convolutional encoder uses finite state machines and that means the finite state diagram will be used to define the internal operation of the encoder. The encoding process involves addition of lock bit i.e. addition of zero or one bit before submitting to the convolutional encoder to make it compatible with the decoding side. Locking binary convolutional encoder is what we call locked convolutional encoder. Locking a binary convolutional encoder does not need changes on the internal structure of the convolutional encoder. Convolutional encoder can be locked by adding either two low bit (zero, zero i.e. 00) or high bit (one, one i.e. 11) after every data bit that will be decoded[6, 7]. Referring to Figure 1, locking by lower locked encoder ignores state three (i.e. $S_3$) and by locking the higher end locked encoder ignores state zero (i.e. $S_0$). Ignoring those states does not change anything, both will work perfectly with the three other remaining states.

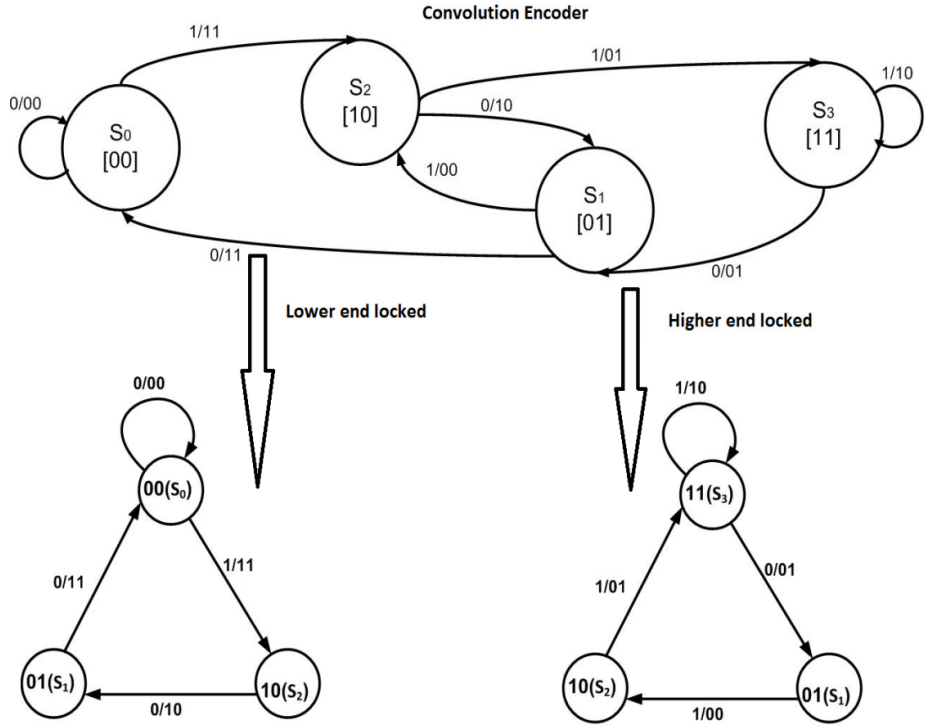

Figure 1: Convolutional Encoder and Locked Convolutional Encoder [6]

## 2.2. Soft Viterbi Decoder

The Viterbi decoder can be either hard or soft decision decoding[22]. In hard decision the received symbol at a sampling instant and quantizer, quantizes the sample value to either '0' or '1'. Simply the hard decision decides whether the received bit is a one or zero by setting the threshold as shown in Figure 2. On the other hand in soft decision the quantizer is a multilevel quantizer[23]. It tries to improve the error correcting capability by utilizing the information concerning the reliability of the received symbol and not threshold setting for this case[24] , as indicated in Figure 3. In Soft decision decoding, the received codewords are compared with all possible codewords and the codewords which give the minimum Euclidean distance is selected. When reading the data, there are sequences of voltage samples corresponding to the parity bits that the transmitter has sent. The decoder is given stream voltage samples and uses that analog information in digitized form using analog to digital conversion in decoding data. Thus, the soft decision decoding improves the decision making process by supplying additional reliability information[24]. The read sequence is a real value and we do not put the threshold because it is Soft decision. It involves add, compare and select to find the minimum metric paths Euclidean distance and select the choice of the minimum metric path called the survivor path[25].The soft decision voltage levels of the received signal at each sampling instant are different as shown in Figure 3. The calculation of the Euclidean distance for the soft decision block is calculated by using the received signal and all possible codewords. You can use either the minim or maximum

International Journal of Computer Science, Engineering and Applications (IJCSEA) Vol. 7, No. 1, February 2017Euclidean distance. The soft decision uses all the voltage levels in this case in making decision. Using Soft decision decoding scheme will improve the performance of the receiver by approximately 2dB when compared to hard decision[14]. If the readsequence $r = m + e$(i.e. read signal 0.7V, 0.4V, 0.2V) where $m$ is a convolutional code sequence and $e$ is an error. The soft Viterbi algorithm examines the trellis diagram for an all zero state to all zero state path whose output sequence is closest to $r$ in terms of Euclidean distance. Finally, it takes the encoder input sequence corresponding to the optimal path to be the most likely message sequence. The output codewords is written to storage media where 0 is written as 'OVolts' and 1 as '1Volts'. The signal corrupts when copied to the storage media and it isread as a distorted data. The soft decision calculates the Euclidean distance between the readsignal and all possible codewords. The decoder then selects the minimum Euclidean distance that matches to the codewords which were written. The soft decision uses all information of the different voltage levels in the process of making decision.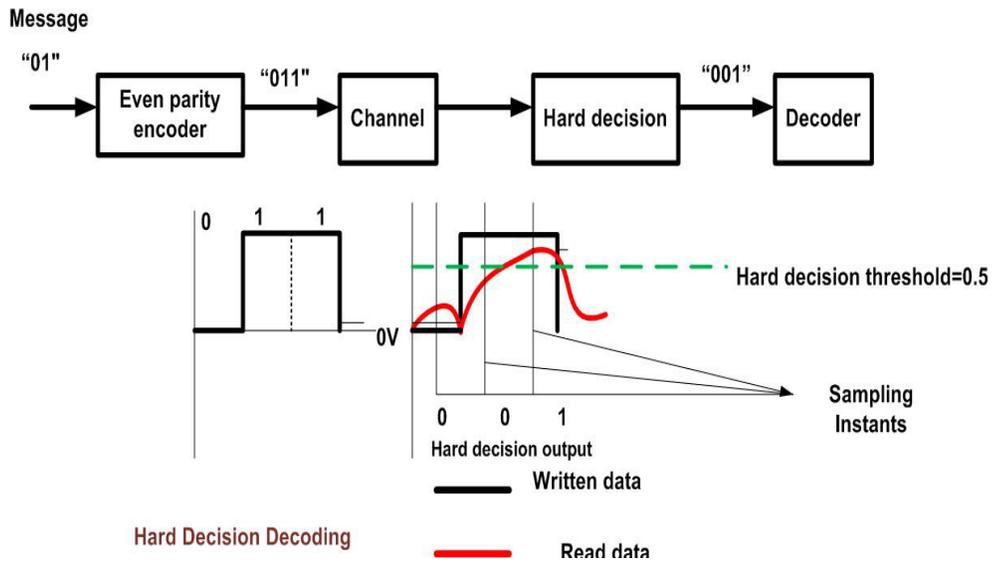

Figure 2: Hard Decision Decoding [14]

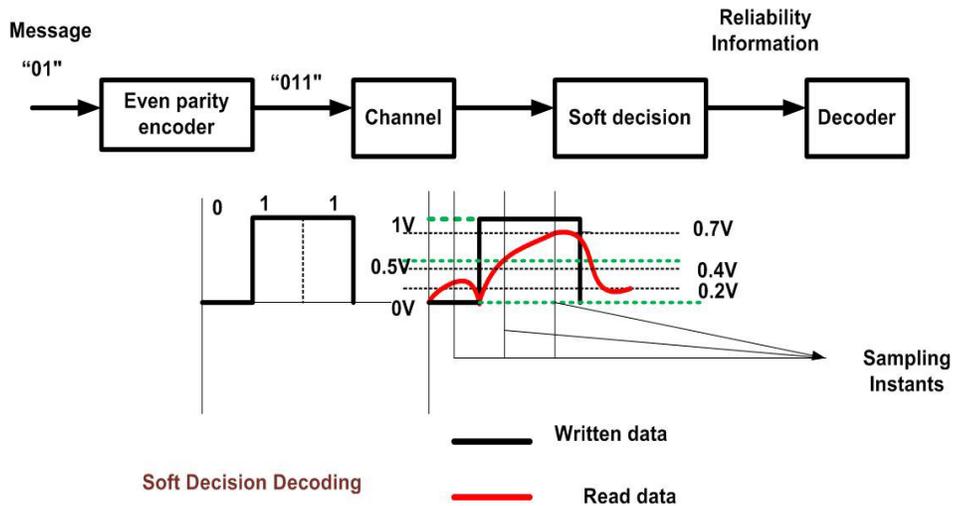

Figure 3: Soft Decision Decoding[14]

Soft decision decoding compares the read codewords with all other possible codewords which give the minimum Euclidean distance. The decoding works on quantized real values because they use a multilevel soft value to represent the output bit.

Simple explanation takes;

Messege => $Ms = (1,0,1,0)$
Message after Modulation $Md = ((1,1), (1,-1), (1,-1), (1,-1))$
The read message $Mr = ((0.7, 0.8), (0.9, -0.7), (-0.7, 0.6), (0.4, -0.8))$
Through the Euclidean distance we can get back to $Ms = (1,0,1,0)$

Figure 4 is the trellis diagram of 1/2 Viterbi decoder with four states $S_0$, $S_1$, $S_2$ and $S_3$. The original codewords written to the storage media devices were $(1, 1), (1, -1), (1,-1), (1,-1)$ and the read codewords were $(0.7, 0.8), (0.9, -0.7), (-0.7, 0.6)$ and $(0.4, -0.8)$. Euclidean distance with its cumulative value for all the paths is at time T. The obtained cumulative distance is compared where the minimum cumulative Euclidian distance indicates the correct path. For example, at time T=0, we have a cumulative distance of 6.13 from $S_0$ to $S_0$, and 0.13 from $S_0$ to $S_1$. In this case, our path will be $S_0$ to $S_1$ because it has the minimum cumulative Euclidian distance. At time T=4, we have a cumulative distance of 5.28 from $S_0$ to $S_0$, 6.88 from $S_0$ to $S_1$, 10.48 from $S_2$ to $S_0$, 3.23 from $S_1$ to $S_2$, 8.88 from $S_2$ to $S_1$, 8.08 from $S_1$ to $S_3$, 14.48 from $S_3$ to $S_2$ and 9.68 from $S_3$ to $S_3$. At this time, the minimum cumulative Euclidian distance is 3.23 which determines our survivor path.

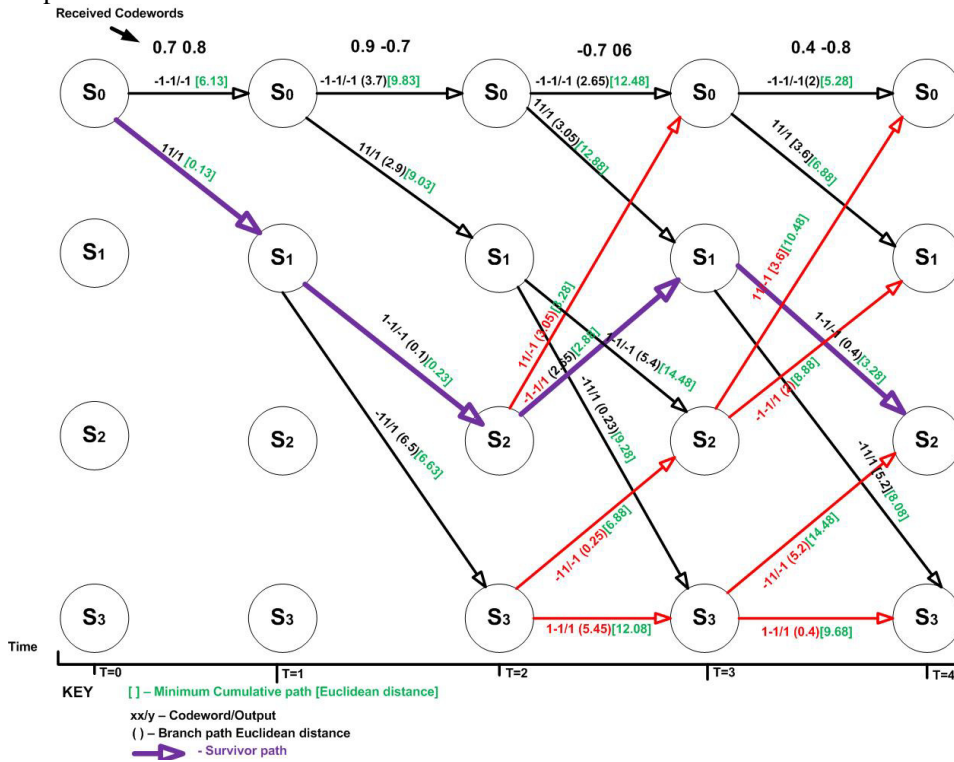

Figure 4: Trellis Diagram Showing Soft Viterbi Algorithm Decoder

## 1.1. Enhanced Soft Viterbi Algorithm Decoder (SVAD-NTCs)

Soft Viterbi Algorithm Decoder that is enhanced with Non Transmittable codewords (NTCs) is used in this design. Viterbi algorithm decoder has special characteristics which enable it to use Non Transmittable Codewords (NTCs) during the reading process [26]. The Non-Transmittable Codewords (NTCs) are the addition of bits to the codewords just before submitting to the Viterbi algorithm decoder in order to improve performance of the decoder. In this design, the NTCs that can be used are either two one-one bits (i.e. 11) for lower end locked encoder or two (–one)-(-one) bits (i.e. -1-1) for higher end locked encoder during the decoding process in soft decision decoding.The behavior of the NTCs show that beyond 6 NTCs there is no significant change on the performance and, this means only 6NTCs will be used. The additional bits (NTCs) make the Enhanced Soft Viterbi Algorithm decoder stable. The added NTCs are removed after the data decoding process.

## 1.2. Proposed Model

Figure 5 is the proposed model for SVAD-NTCs. The data are randomly generated by a binary generator and then sent to the locked convolutional encoder. Then additional of lock bit i.e. zero-zero in the locked convolutional encoder before data are written into the storage media. Data source encoded by Locked convolutional encoder so that it can be possible to decode them through Soft Viterbi decoder enhanced with NTCs. The encoded data are modulated and written to the storage media devices. In the storage media, it is where data can be distorted or corrupted thus making us fail to retrieve the stored data. To be able to retrieve the corrupted data you have to use powerful algorithm which will be able to recover the corrupted data. In this case the stored data will be retrieved by using Soft Viterbi Algorithm decoder enhanced with NTCs (SVAD-NTCs) to get the original data. In reading data from the storage media the data are demodulated then, addition bits are added to the demodulated data before they are sent to SVAD-NTCs. These addition bits are the Non-Transmittable Codewords(NTCs) which will be negative one-negative one. For this case,six NTCs will be added to the read codewords. After decoding using SVAD-NTCs addition bits both lock bits and NTCs are removed to get the original data.

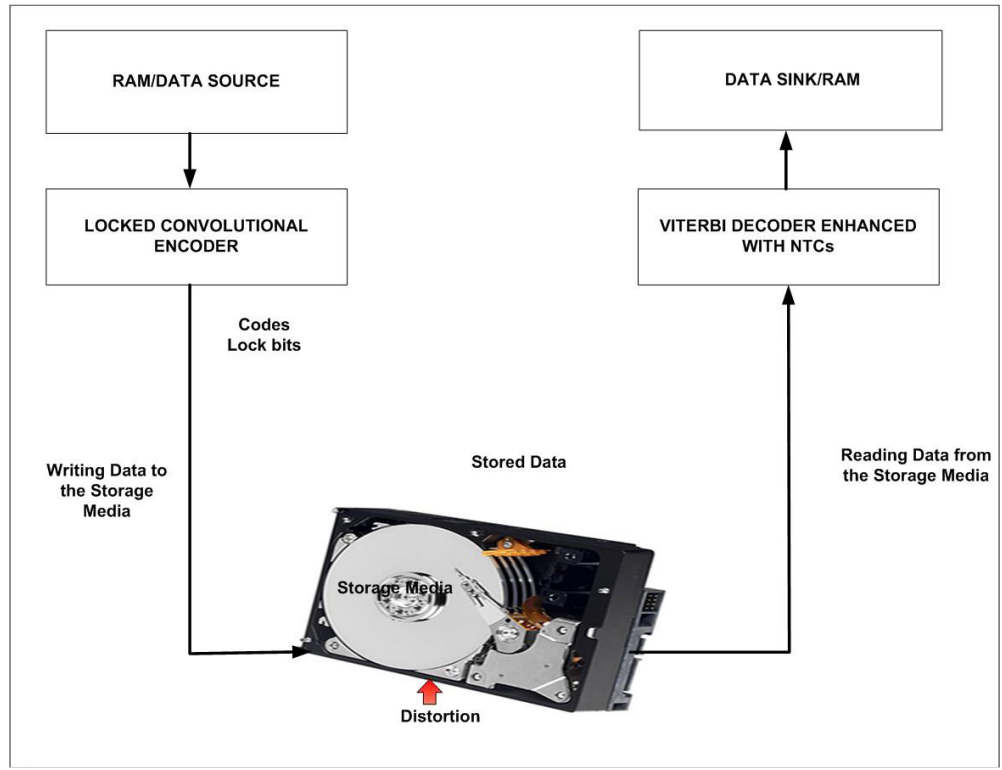

Figure 5: Proposed Model

## 3. SIMULATION

One million data bits were randomly generated by random function, where the obtained set is duplicated into two copies. The first copy is submitted to the locked convolutional encoder for lock bit addition and encoding. The second copy is sent to the Reed Solomon encoder. Both data are then modulated by Binary Phase Shift Keying (BPSK).Both sets are subjected to Additive White Gaussian Noise (AWGN) that mimic the storage media distortion. After distortion, both sets are demodulated and then each set is submitted to its corresponding decoder. Data coming from the Reed Solomon decoder are submitted to Reed Solomon decoder for decoding to obtain the final set which is compared to the original data set. Six NTCs are added to data coming from the Locked Convolutional encoder and then submitted to the enhanced Soft Viterbi Algorithm decoder. After the decoding process, both lock bit and NTCs are removed leaving behind a set of data to be compared to the original data set. Eventually a set with minimum divergence from the original set is identified. Figure 6 is the simulation block diagram.

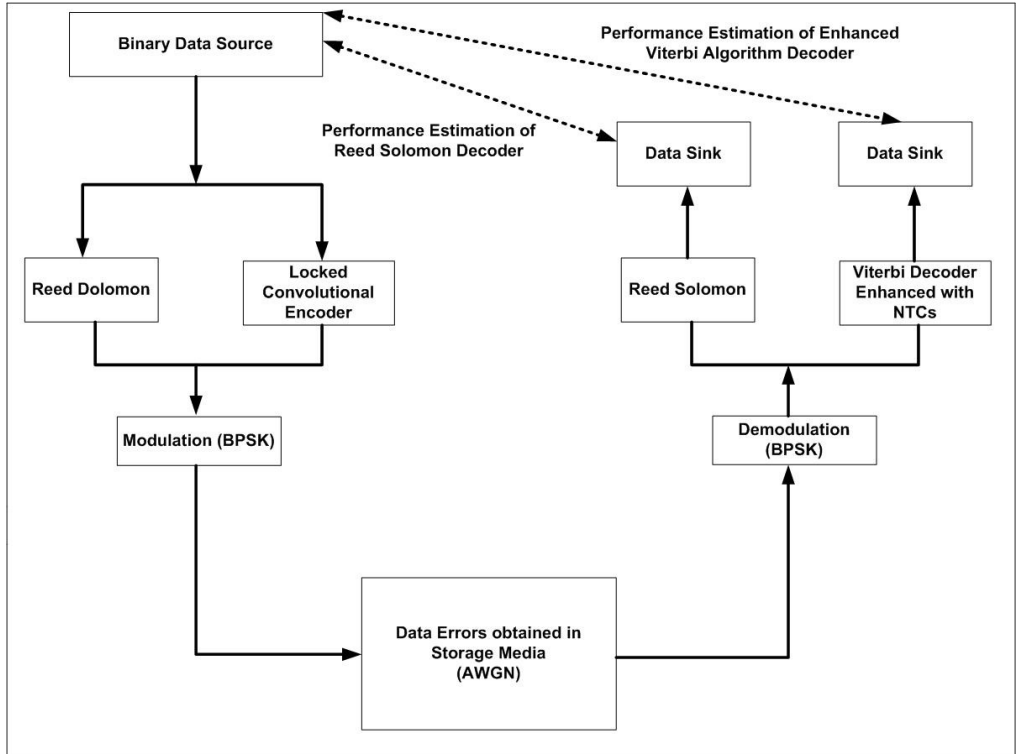

Figure 6: Simulation Block Diagram

## 4. DISCUSSION

The collected data from both SVAD-NTCs and Reed Solomon algorithm were compared with the original data set. Table 1 shows the comparison of the residual error whereas signal to noise ration 6 and above the SVAD-NTCs was able to correct almost all errors from the storage media. It was able to correct the error by 99.9966%.As Eb/No ration in dB becomes small, that means there are more errors and the residual is high. Both algorithms showed there strength at this point which was important to see their strength in correcting the error. For example, at Eb/No equal to 1, SVAD-NTCs was able to correct the errors and remain with only 13,973 out of one million, and the Reed Solomon corrected and remained with 78,933 out of one million which is around 6 times better than the Read Solomon. Generally, the results show that SVAD-NTCs design has reduced the total residual errors from 216,878 of Reed Solomon to 23,900. Results show the improved ability in correcting the residual error. Again, this results indicate that SVAD-NTCs increases performance and reliability of the storage media and we recommend the researchers to work on this technique and implementation on storage media devices should be considered.

Table 1: Residual Error for SVAD-NTCs and Reed Solomon Algorithm

| Eb/No | RS Residual errors after decoding | SVAD-NTCs Residual errors after decoding |
|---|---|---|
| 1 | 78933 | 13972 |
| 2 | 56335 | 6352 |
| 3 | 37208 | 2548 |
| 4 | 22744 | 786 |
| 5 | 12274 | 197 |

| 6 | 6035 | 41 |
| 7 | 2350 | 4 |
| 8 | 760 | 0 |
| 9 | 212 | 0 |
| 10 | 22 | 0 |
| 11 | 5 | 0 |
| Total | 216,878 | 23900 |

Figure 6compares error correction strength of RS Algorithm decoder and SVAD-NTCs decoder. Where at signal to noise ration 1 the graph shows that the residual errors are below 14,000 for the SVAD-NTCs and below 80,000 for Reed Solomon decoder. At signal to noise ration 4, SVAD-NTCs show that they can correct almost all the residual errors while at the same point the RS still remains with more than 22,000 residual errors. At signal to noise ration 8 RS algorithm decoder corrects almost all the errors which are 4dB less than the SVAD-NTCs can perform. As the distortion decreases the signal to noise ratio increases. But our interest is when the signal to noise ratio is small. At this point, the SVAD-NTCs perform better than the RS.

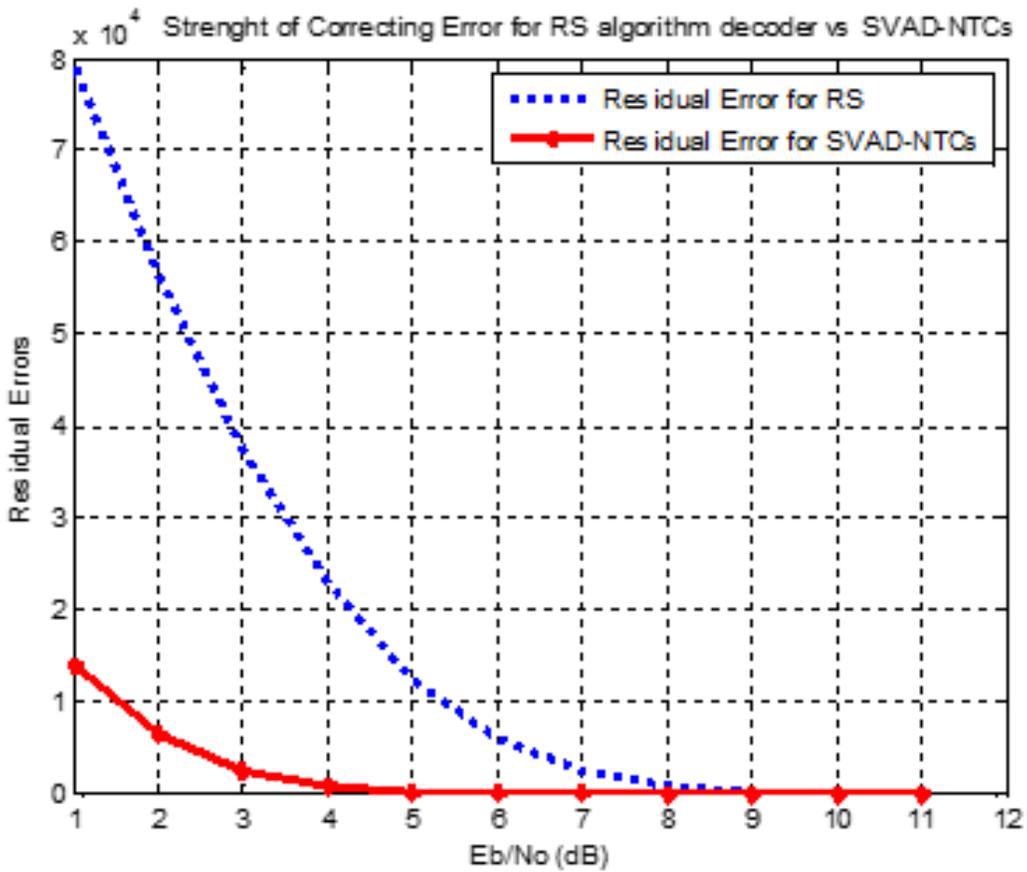

Figure 7: Errors Correction Strength for RS Vs SVAD-NTCs Decoder

International Journal of Computer Science, Engineering and Applications (IJCSEA) Vol. 7, No. 1, February 2017

## 5. CONCLUSION

This paper implemented design of the Soft Viterbi Algorithm Decoder enhanced with Non-Transmittable Codewords which is proposed tobe used in the storage media. The locked convolutional encoder and SVAD-NTCs wereimplemented in design.BPSK and AWGN were used in Matlab simulation to evaluate the parameter of the model. The simulation results show that SVAD-NTCsare more powerful in error correction which is 6times better than Reed Solomon algorithm in the residual errors. At 4dB the SVAD-NTCs corrects almost all the residual errors while the RS does it at 8dB. This means that the SVAD-NTCs is 4times better than the RS algorithm. This indicates that the design can improve storage media reliability. Therefore, it is recommended that further researchshould be on the proposed algorithm (SVAD-NTCs) to increase the storage media reliability

International Journal of Computer Science, Engineering and Applications (IJCSEA) Vol. 7, No. 1, February 2017

## AUTHORS

**Kilavo Hassan:** Currently he is the Ph.D. student at Nelson Mandela African Institution of Science and Technology in the School of Computational and Communication Science and Engineering majoring Telecommunication Engineering. He received his B.Sc. Computer Science at the University of Dar es Salaam, Tanzania and His M.Sc. in Information and Communication Science Engineering majoring Telecommunication Engineering at the Nelson Mandela African Institution of Science and Technology Arusha Tanzania. His research interest includes Forward Error Correction, Embedded System and Automation System.
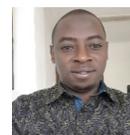

**Kisangiri Michael**, received his Ph.D. degree in Telecommunication Engineering (2008) and Master of Science in Telecommunication Engineering (2002), both from Wroclaw University of Technology - Poland. Since then he has been working as academician with several universities. Currently he works with Nelson Mandela African Institution of Science and Technology located in Arusha, Tanzania. His research interests include evolutionary computation in Telecommunication Networks, Antenna design and Triangular mesh modeling.
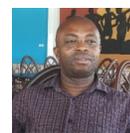

**Salehe I. Mrutu** received his B.Sc. and M.Sc. degrees in Computer Science from The International University of Africa in 2003 and the University of Gezira in 2006, both Universities from the Republic of Sudan. He received his Ph.D.in Information and Communication Science and Engineering in 2014 at Nelson Mandela African Institution of Science and Technology (NM-AIST) in Arusha. Currently Dr. Mrutu is serving as lecturer at the University ofDodoma (UDOM) under the College of Informatics and Virtual Education. His research interests include Forward Error Correction (FEC) Codes, algorithms analysis and designing, quality-of-service provisioning and resource management for multimedia communications networks.
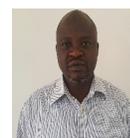